# Terrestrial Orbit-Spin Coupling Torque Episodes in Late 2020

James H. Shirley

*Jet Propulsion Laboratory, California Institute of Technology, Pasadena, CA, USA*

5 October 2020

**Abstract:** Orbit-spin coupling torques on the Earth in November 2020 are larger than at any other time between 2000 and 2050. This affords an opportunity to observe the terrestrial atmospheric response to the putative torque in near real time.

Figure 1 illustrates the variability with time of orbit-spin coupling torques on the Earth in 2020. Fig. 1 is annotated with a number of key dates. The peak dates of three future torque episodes are indicated. The torques peak near 12 October, 8 November, and 7 December 2020. The dates of the minima separating the episodes are also of interest. Earth recently passed through a near-zero-torque interval at the beginning of October. Subsequent minima are found on days 296, 326, and 355 of 2020, corresponding to 23 October, 21 November, and 21 December, respectively.

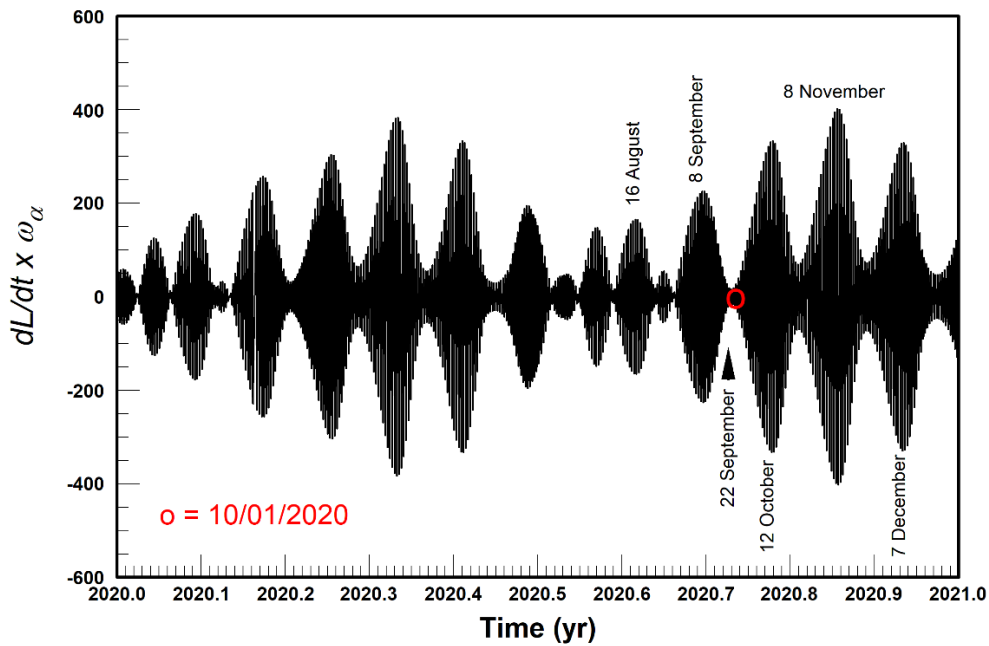

**Figure 1**. Diurnal and seasonal variability of the magnitude of the orbit-spin coupling torque for a location on Earth's equator (time step 6 hours). See Methods (below) for calculations.

The orbit-spin coupling hypothesis [1] predicts that an intensification, or powering up, of the large-scale circulation of the atmosphere will occur at times when the torques are large. We thus expect that momentum will be deposited, within the terrestrial atmosphere, as a consequence of the torque, during each of the upcoming torque episodes of Fig. 1. Experience gained through extensive simulation of the Mars atmosphere suggests that a strengthening of meridional overturning circulations may be expected, together with a strengthening of pressure gradients for mid-latitudes cyclones and anticyclones. Stronger variability of local weather conditions is an expected outcome with a more vigorous large-scale circulation. In the absence of numerical modeling results for this subject body and this time period, we cannot at present more accurately specify the changes in atmospheric conditions and behaviors that may accompany the anticipated spinning-up of the atmospheric circulation during these three episodes.

The year 2020 has been one of remarkable extremes of weather and climate. It is beyond the scope of this short note to attempt to correlate past weather events with all of the oscillations displayed in Fig. 1. However, we have added three additional key dates to Fig. 1 that identify two recent torque episode peaks, together with one relaxation-phase date. Consideration of these episodes yields added perspective on possible outcomes during the three future torque episodes of Fig. 1.

The peak of the mid-August torque episode of Fig. 1 corresponds in time to an outbreak of >12,000 lightning strikes in Northern California. Forest fires from that episode are still burning. (The excess energy of forced atmospheric motions is largely dissipated by frictional processes [2]).

The peak of the early September torque episode corresponds in time to the occurrence of the early-season snowstorm that brought extreme temperature changes in Boulder and Denver.

Each of these examples describes phenomena that may result from or be associated with a powering-up of the large-scale circulation on short timescales. The third example relates to the opposite condition, when the torques are much reduced. Experience at Mars has shown that the spinning-down of a previously intensified circulation may likewise give rise to distinctive weather events and conditions [2]. The arrow symbol at 22 September 2020 corresponds to the stalling of tropical storm Beta over the U. S. Gulf coast. This storm dumped ~10 inches of rain in Southeast Texas and produced flooding in Texas, Arkansas, and Louisiana. The relaxation phase of the torque "cycle" is characterized by much reduced deposition of momentum within the atmosphere. At such times, the spun-up atmosphere may "take a breather," plausibly accounting for phenomena such as the stalling-out of tropical storm Beta.

We will assess the applicability and accuracy of the above predictions after the highlighted three-month period is completed. Future work will more fully characterize the nature of the terrestrial atmospheric response to the orbit-spin coupling torques. Possible lags in the response are of interest. Quantitative analyses may be formulated and are encouraged.

**Background:**

The orbit-spin coupling hypothesis [1] predicts driven cycles of intensification and relaxation of planetary atmospheric circulations. The coupling, given by expression (1), takes the form of a reversing torque [1-4] with axis lying in the equatorial plane of the subject body;

$$- c\, (\dot{\mathbf{L}} \times \boldsymbol{\omega}_a) \times \mathbf{r}. \qquad (1)$$

Here $\dot{\mathbf{L}}$ (or $d\mathbf{L}/dt$) represents the time rate of change of the orbital angular momentum of the subject body with respect to the solar system center of mass (or barycenter), while the axial rotation of the subject body (with respect to the same inertial coordinate system) is represented by the angular velocity vector $\boldsymbol{\omega}_a$. $\mathbf{r}$ denotes a position vector in a rotating Cartesian body-fixed coordinate system, while $c$ is a scalar coupling efficiency coefficient, which is constrained by observations of planetary motions to be quite small [1, 3-5]. Expression (1) describes a global acceleration field, with units of m s$^{-2}$, wherein the accelerations everywhere lie in directions tangential to a spherical surface.

The orbit-spin coupling hypothesis has been remarkably successful in explaining the intermittent occurrence of global dust storms on Mars [3-5]. Atmospheric general circulation model simulations [4, 5] reveal that an intermittent strengthening and weakening of meridional overturning circulations is a characteristic feature of the mechanism. The predicted intensification was observed by spacecraft observations at the start of the Martian global dust storm of 2018 [6]. Adding orbit-spin coupling accelerations to Mars GCMs significantly improves the agreement between atmospheric simulations and observations [4-5, 7].

Recent work [7] indicates that the orbit-spin coupling torques on the Earth are significantly larger than those on Mars. The terrestrial torques, in addition, cycle much more rapidly, due to the presence of Earth's Moon. (The nearby Moon gives rise to significant time variability of the Earth's orbital angular momentum with respect to the solar system barycenter). We have thus begun to investigate the question of the possible response of the Earth atmosphere to the deterministic orbit-spin coupling torques given by expression (1).

**Methods**

In prior work we have presented curves representing the time rate of change of the orbital angular momentum (with respect to the solar system barycenter) ($d\mathbf{L}/dt$) in order to characterize the orbit-spin coupling "forcing function." No information regarding the quasi-diurnal cycle of the acceleration at any specific location is provided by this approach. In Fig. 1 of this note, we have gone one step farther, forming the cross product ($\dot{\mathbf{L}} \times \boldsymbol{\omega}_a$) as a function of time. This identifies a vector lying in the equatorial plane of the subject body. While the direction in inertial space of this vector does not greatly change over short periods, its direction within the conventional body-fixed Cartesian coordinate frame does cycle (approximately over the sidereal period) as a consequence of Earth's rotational motion. The curve in Fig. 1 is obtained by sampling the quasi-sinusoidal variation of the x or y component of the ($\dot{\mathbf{L}} \times \boldsymbol{\omega}_a$) vector at 6 hour intervals. We emphasize that this approach serves mainly to reveal the variability of the magnitude of the torque. Focused studies will require a more sophisticated approach.

Methods for obtaining the angular momentum of Earth (or any other solar system body) with respect to the solar system barycenter are described in [1-3]. To obtain units of acceleration, the values on the y-axis of Fig. 1 must be multiplied by two factors: the value (in meters) of $\mathbf{r}$, the 3-component position vector for a given location; and $c$, the coupling efficiency coefficient. The coefficient value has not yet been determined for the case of the terrestrial atmosphere. An optimized $c$ value obtained for the Mars atmosphere [4] was $c=5.0 \times 10^{-13}$.

**Acknowledgements**

Portions of this work were performed at the Jet Propulsion Laboratory, California Institute of Technology, under a contract from NASA. Copyright 2020, California Institute of Technology. Government sponsorship acknowledged.